# Information content of and the ability to reconstruct dichroic X-ray tomography and laminography

**MATTHEW A. MARCUS**

*Advanced Light Source, Lawrence Berkeley Laboratory, 1 Cyclotron Road, Berkeley CA 94720 USA*
*\*mamarcus@lbl.gov*

**Abstract:** Dichroic tomography is a 3D imaging technique in which the polarization of the incident beam is used to induce contrast due to the magnetization or orientation of a sample. The aim is to reconstruct not only the optical density but the dichroism of the sample. The theory of dichroic tomographic and laminographic imaging in the parallel-beam case is discussed as well as the problem of reconstruction of the sample's optical properties. The set of projections resulting from a single tomographic/laminographic measurement is not sufficient to reconstruct the magnetic moment for magnetic circular dichroism unless additional constraints are applied or data are taken at two or more tilt angles. For linear dichroism, three polarizations at a common tilt angle are insufficient for unconstrained reconstruction. However, if one of the measurements is done at a different tilt angle than the other, or the measurements are done at a common polarization but at three distinct tilt angles, then there is enough information to reconstruct without constraints. Possible means of applying constraints are discussed. Furthermore, it is shown that for linear dichroism, the basic assumption that the absorption through a ray path is the integral of the absorption coefficient, defined on the volume of the sample, along the ray path, is not correct when dichroism or birefringence is strong. This assumption is fundamental to tomographic methods.

## 1. Introduction

Throughout the history of their use, X-rays have been used to make two-dimensional, and later, three-dimensional images encoding the distribution of elements or chemical species, which are expressed as scalar quantities. However, in many cases, one is interested either in magnetic moments (vector) or crystal orientation (tensor). The imaging of magnetic structures has been of major importance in the study of spintronic and other magnetic materials, and many methods have been developed for imaging magnetization in two dimensions. One important class of such methods makes use of the phenomenon of magnetic circular dichroism (MCD), in which the X-ray absorbance of a material differs between left- and right-circular polarization by an amount depending on the local magnetization[1, 2]. This effect is readily used to create two-dimensional images in various sorts of X-ray microscopes such as ptychographs[3] or photoelectron emission microscopes[4]. Note that these two papers are only examples; many more exist in the literature. However, only recently has there been work in which the magnetic order, as a three-dimensional vector, been imaged in three dimensions[5]. It should be noted, however, that ultrasound tomographic reconstruction of flow fields in fluids is mathematically equivalent to MCD tomography and was theoretically investigated before the X-ray version was accomplished[6].

Another important class of materials is crystals, in which the orientation is an important part of the microstructure and plays an important role in understanding the properties and functions of crystalline materials. For instance, teeth are made of apatite, which is not a particularly strong or durable material, but the nanostructure of enamel makes them wear-resistant enough to last a lifetime[7]. While diffractive methods such as electron backscatter diffraction and X-ray diffraction tomography can provide very useful information in two and three dimensions, they have limitations in sample thickness (electron diffraction) or resolution on the nanoscale or fail if the material is not perfectly crystalline, as is often the case for biominerals. What is wanted is a method which can image the orientation of a crystalline material, over cubic microns of volume, with sub-micron or nanoscale resolution. Although methods such as PEEM-PIC mapping[8] and dichroic ptychography [9-11] provide the requisite resolution and sampling area in two dimensions, it can be very difficult to image orientations in three dimensions.

Dichroic tomography is an extension of the well-known technique of X-ray computed tomography, in which a sample is rotated in an X-ray beam and, at each of a set of rotation angles, a projection image (radiograph) is captured. This set of images is processed to produce a 3D reconstruction of the absorption density of the sample. In dichroic tomography, the incident beam is polarized, and so additional contrast is provided due to the crystal orientation (linear dichroism) or magnetization (circular). Magnetic linear dichroism, as observed in antiferromagnets, is equivalent to the linear dichroism seen in uniaxial crystals (see, for example [12]). In this case, orientation is replaced by antiferromagnetic order parameter with no change in the formalism and so is covered by discussion of XLD.

The purpose of this paper is to discuss the theory of how the properties of the sample map onto the projected images, and what measurements would be required in order to make it possible to reconstruct not only an overall absorption at each voxel, but the anisotropy. While the development is done in the context of reconstruction by filtered back-projection, the purpose is not to propose a reconstruction algorithm but to ask what would be required to make such an algorithm possible. This leads to considerations such as:
1. What assumptions must we make and under what conditions are they valid?
2. How many polarizations need we use? Can we get the whole set of parameters (three components of magnetization, or full orientation and magnitude of linear dichroism)?
3. What is the difference between computed tomography geometry (CT, rotation axis perpendicular to beam) and computed laminography (CL, rotation axis at an angle to the beam other than 90º)?

The next section goes over a number of definitions and discuss the framework and formalism used for the math. Then I'll specify some assumptions being made and what they imply, including a review of filtered back-projection, after which comes the actual derivations.

This paper was originally written in the context of soft X-ray scanning transmission X-ray microscopy (STXM), but the results should apply to hard X-rays and to full-field imaging methods.

In the course of the development, some results will be derived which were previously known. These serve as examples and checks on the mathematical methods.

## 2. Preliminaries
This section is where definitions and formalism are laid out.

### 2.1 Assumptions
First, some assumptions:

> *A1: The absorption through the material, from source to detector, may be described by Beer's law applied to each voxel along the line of sight. Thus, there is such a thing as a volume absorbance, which has an angular and polarization dependence for a dichroic material.*

The result of a normal tomographic reconstruction is the volume absorbance of the set of voxels defined by positions in the sample, $\mu(x, y, z)$. Thus, the beam intensity after passing through a small thickness $dl$ of material at this position is attenuated by a factor of $1 - \mu(x, y, z)dl$. For an isotropic material, this quantity is independent of beam direction or polarization, an assumption implicit in all standard tomographic reconstruction algorithms. This assumption is fundamental to most, if not all tomographic reconstruction algorithms.

An important way in which this assumption is violated is if the polarization of the beam changes as the beam propagates through the material. As an extreme example, consider a material which is so strongly dichroic that a small thickness of it acts like a polarizer. If the incident polarization is vertical but a grain has a vertical absorbing axis so it acts like a horizontal polarizer, the transmitted intensity is low. If, however, a grain oriented at 45° is placed before the original grain, the interposed grain turns the polarization, and now up to 1/4 of the incident beam goes through. If the order of the grains is switched, the transmission drops again. This shows that a simple line-integral treatment of absorption does not work in strong dichroism conditions. The transmission is not a scaler, but a tensor described not by a line integral but by a product of non-commuting matrices. This problem does not occur in the MCD case because the polarization remains left or right circular as the beam passes through the sample[13]. The propagation of X-rays through a linearly-dichroic material is described in a number of places including [9].

What can be done to make A1 true or nearly true? One way is to operate in a regime of weak dichroism and birefringence. This of course can cause trouble with noise and artefacts, but at least guarantees that the polarization remains nearly unchanged through the sample. Note that operating at an energy below or near the dichroic peak, even if the dichroism of absorption is small, may not be appropriate because the sample may still be birefringent enough to act like a stack of waveplates and scramble the polarization. In analysis, it may be possible to adopt some kind of tensor formalism, especially if real and imaginary parts of the transmission are known as would be the case if the projections are derived from ptychography. It may also be possible to come up with an algorithm incorporating a multislice model as the forward model in an iterative loop.

The next assumption refers to the use of ptychography:

> *A2: Ptychography is just a high-resolution form of STXM. The result of ptychographic reconstruction at a given beam direction and polarization is an image which reduces to an optical density at each pixel which is given by a line integral of the volume absorbance along the beam direction. The volume absorbance at a voxel is that of a bulk piece of material of the same composition and orientation as that of the material of the voxel. Ptychography actually provides a complex transmission, implying a complex absorbance. For instance, the phase shift through the sample can be imaged instead of the absorption, but the math is the same as for the real case. The complex transmission can be written as $\exp(-OD + i\phi)$ where OD is the optical density and $\phi$ is the phase shift.*

In the presence of linear dichroism, this assumption should be applied carefully. The reconstruction of the transmitted field in principle needs to take into account the tensor nature of the index of refraction [9]. However, once the reconstruction is done, the squared magnitude of the transmitted field plays the role that transmission does in STXM or the phase shift that of absorption in STXM (see for example[14]).

Now comes an assumption which may not be necessary but makes the math simpler:

> *A3. No missing angles. It is assumed possible to rotate the sample through a full 360° over as many angles as necessary for reconstruction..*

The next assumption involves terms which will be defined below, so readers who don't know what back-projection is should read to that section and come back. This assumption is:

*A4. The back-projection derived from all the projections uniquely maps onto the projections and therefore contains all the information in the projections, which is all the information there is.*

This is actually a theorem, rather than just an assumption. The proof is as follows: The contribution to the back-projection from a single pixel of any of the projections is a line in real space, extending along the beam direction corresponding to that projection. Thus, its Fourier transform is a sheet in Fourier space, oriented perpendicular to the beam direction and passing through the origin. Now consider the Fourier transform of a single back-projection, which is the contribution to the back-projection from a single rotation. Since the lines from all points on the detector are parallel, they contribute to the same sheet in Fourier space, which now contains the Fourier transform of the projection. Since all the sheets are unique, it follows that the Fourier transform of the back-projection can be unfolded to give the Fourier transforms of each projection, which can be inverted to yield the set of projections. For a finite number of rotations, the sheets intersect each other at a set of measure zero compared with any one sheet, so a point on a sheet corresponds uniquely to a point on a projection's Fourier transform.

Finally, an assumption which provides an important simplification for dichroic CT/CL:

*A5: The incident beam is collimated, so its direction, and hence the polarization vector is the same for all points in the sample.*

Thus, we do not consider cone-beam imaging or imaging with a focused probe for which the thickness of the sample is significant compared with the depth of focus.

*2.2 Definitions*

Next, a number of definitions:

All unit vectors will be written with hatted symbols, e.g. $\hat{n}$ (beam direction), while non-normalized vectors such as a position $\vec{X}$ will be written with over-arrows.

The method consists of rotating the sample in the beam while taking radiographs. Each radiograph is processed assuming Beer's Law to produce the projected absorption along the beam path. These are called *projections*, and are parameterized by the rotation angle $\gamma$ and any other variables such as polarization. For phase contrast (e.g. from ptychography), the projected phase shift is the quantity given in the projections.

If the rotation axis is perpendicular to the beam, then we have *computed tomography (CT)* geometry and each projection may be regarded as a stack of projections of 2D slices. Otherwise, we have *computed laminography (CL)* geometry and no such slicing is possible – the whole volume must be reconstructed at once. While CL is more complex than CT, and as we shall see, has an intrinsic limitation due to missing information, it is often the only way to work with samples that are sheet-like in geometry, as CT will result in projections in which the sample is probed on-edge. This case results in ranges of rotation angles in which the data are useless because the sample is effectively too thick to see through. Also, for circular dichroism, CL offers the potential of detecting all three components of magnetization, while CT misses the component along the rotation axis.

Two coordinate systems will be used. One is the *lab coordinate system* (or reference frame), in which the beam comes from a fixed direction, the detector is in a fixed position, and the sample rotates. The other is the *sample coordinate system* (or reference frame) in which the sample is fixed and the source and detector pirouette around it. This is the coordinate system in which we want the eventual reconstructions to be defined.

It is assumed (A1) that the absorption through the sample (a projection) is described for each angle and voxel by a local *effective absorption coefficient*

$$r(\vec{X};\gamma,\chi) = \ln(I_0/I) = \int dl\, \mu(\vec{x}(l);\gamma,\chi) \tag{1}$$

where $l$ is the length along a path extending to a given point on the detector, $\vec{x}(l)$ is the position in the sample (in its own coordinate system) corresponding to $l$ and the position on the detector $\vec{X}$, $\gamma$ is the rotation angle and $\chi$ is the polarization, specified either as an angle with respect to the horizontal (in the lab coordinate system) or a handedness of circular polarization. The phenomena we're probing are *MCD* (magnetic circular dichroism) or *XLD* (X-ray linear dichroism). XLD may come from crystalline anisotropy, in which case the objective is to solve for the crystal orientation at each voxel, or the second-order linear dichroism of an antiferromagnet, in which case the goal is to solve for the order parameter.

The method of reconstruction we will consider involves the creation of back-projections, made by adding up single back-projections for each rotation angle. A *single back-projection*, parameterized by the rotation angle is made by projecting a projection back through the sample to create a 3D density:

$$p(\vec{x};\gamma,\chi) = \int r(\vec{X} - l\hat{n};\gamma,\chi)dl \qquad (2)$$

where $\vec{x}$ is a position in the sample volume, $r$ is the projection at a given rotation angle, $\hat{n}$ is the beam direction in the sample reference frame, and $l$ is the length along the beam from the sample voxel $\vec{x}$ to the plane of the detector.

The *back-projection* is the sum of the set of single back-projections, and looks like the object with a blurry halo around it. This halo, at least in the isotropic case, can be removed by a Fourier filtering procedure, resulting in a *filtered back-projection* as an approximation of the original sample.

To get from a set of projections or a back-projection to the unknown parameters of the sample, we must use some sort of mathematical method which essentially does a large linear-algebra problem. Reconstruction by filtered back-projection (FBP) is the core of one family of methods used for isotropic samples. The other major family of methods considers the set of local absorptions per pixel as a large vector of unknowns, which is related by a linear matrix operation to the set of projections, considered as a larger vector of observations. This is essentially a linear least-squares problem, but the response matrix is huge. The vector of unknowns has as many elements as there are voxels and the vector of observations has a length equal to the product of the number of angles and the number of pixels on the detector. Thus, for a small case of a 200x200x200-voxel sample volume, 360 rotations and a 200x200 pixel detector, we have $8 \times 10^6$ unknowns and $1.44 \times 10^7$ observations, so the matrix would have $1.16 \times 10^{14}$ elements. Even on a supercomputer cluster, it would be impractical to store and invert such a matrix, even though it is sparse. Instead, *iterative least-squares methods* such as SIRT (Simultaneous Iterative Reconstruction Technique[15]) are used. Although these methods are better in some senses than FBP methods and have largely taken over the field, the argument above shows that back-projections contain all the information there is, so FBP methods can be used to investigate the conditions under which there is enough information to reconstruct anisotropic samples.

We will see that the local absorption coefficient varies with rotation angle according to a trigonometric series including sines and cosines of the rotation angle and twice the rotation angle. If one wants the coefficients in such an expansion, one way to do it is to average the function with a weight of $\sin\gamma, \cos\gamma, \sin 2\gamma, \text{or } \cos 2\gamma$. Similarly, the projections $p(\vec{x},\gamma)$ will include trig-function variation, so one can get all the available information by taking averages over $\gamma$ such as $\langle p(\vec{x},\gamma)\cos\gamma \rangle$ with the angle brackets denoting an angular average over rotation angle $\gamma$. These shall be denoted as *weighted back-projections*, and the trig-function coefficients of $\mu(\vec{x},\gamma)$ as *moments*. For some of the analysis, it will be more convenient to expand in terms of $\exp(im\gamma)$, with *m* ranging from -

1 to 1 (MCD) and -2 to 2 (XLD). This is, of course, equivalent to an expansion in sines and cosines.

Much of the mathematical work was done using Mathematica[16]. In particular, the MatrixRank[ ] and Eigenvalues[ ] functions were used in order to evaluate the number of independent equations represented in some linear systems.

## 3. Dichroism Framework

At this point, we can start to create the framework in which we work. First, I will define some geometry. This is where assumption A5 (beam direction is everywhere the same) comes in. Figure 1 shows the assumed geometry, which shows the general CL case and the notation for the relevant angles. This geometry was originally considered for a tomographic electron-microscope sample holder, but is general enough to be applied to other mechanical arrangements.

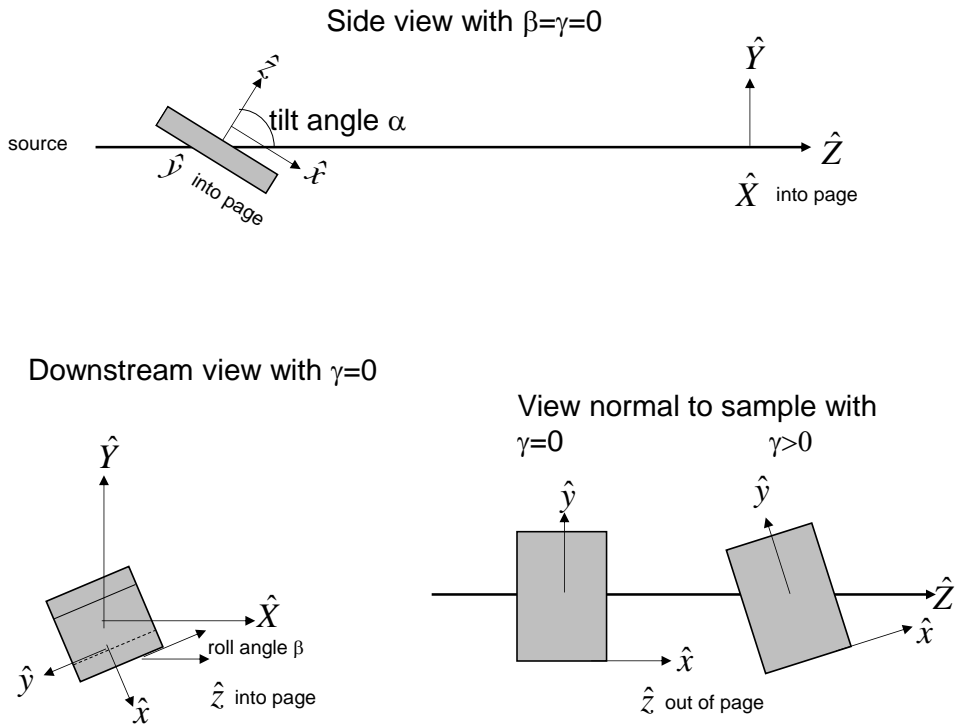

Fig. 1. Geometry of the measurement. The unit vectors $\hat{x}, \hat{y}, \hat{z}$ define the sample coordinate system with $\hat{z}$ the rotation axis. Lab coordinate axes are $\hat{X}, \hat{Y}, \hat{Z}$ with $\hat{Z}$ being the beam direction.

The three angles provided by the mechanics are $\alpha$, the angle between the rotation axis and the beam, so $\alpha = 90°$ is CT mode., $\beta$, a roll angle about the beam axis, and $\gamma$ the rotation angle, which is the only angle that varies between projections. Further, the polarization in XLD measurements is defined in terms of $\chi$, the angle between the polarization vector and the horizontal ($\hat{X}$). The entire sample holder rotates about the $X$ axis

($\alpha$), the plate holding the grid wags up and down by angle $\beta$, and the sample, idealized as a thin object such as a TEM grid, rotates in its own plane by angle $\gamma$. The rotation specified by $\beta$ is equivalent to rolling the sample about the beam axis. This action can be mimicked by rolling the polarization and detector about the beam axis, so $\beta$ has no non-trivial effect. Thus, we will take it to be 0 in the following. Now we can start to assemble the ($\hat{x}\hat{y}\hat{z}$) orthonormal triad in lab coordinates. Start with $\alpha = 90°; \beta = \gamma = 0$. Inspection of Figure 1 shows that the ($\hat{x}\hat{y}\hat{z}$) triad in lab coordinates is:

$$\begin{pmatrix} \hat{x} \\ \hat{y} \\ \hat{z} \end{pmatrix}\bigg|_{90°,0,0} = \begin{pmatrix} 0 & 0 & 1 \\ 1 & 0 & 0 \\ 0 & 1 & 0 \end{pmatrix}. \quad (3)$$

Here and in the next few equations, the pile on the LHS is a pile of row vectors, so is a matrix. The determinant of this matrix is 1, showing that the parity of the sample coordinate system as illustrated in Figure 1 is the same as that of the lab system. The first rotation to apply is $\alpha$, which differs from the other two in that it is done in the lab system, since the whole holder moves. This is a rotation about $X$, which mixes $\hat{Y}$ and $\hat{Z}$, so we have

$$\begin{pmatrix} \hat{x} \\ \hat{y} \\ \hat{z} \end{pmatrix}\bigg|_{\alpha,0,0} = \begin{pmatrix} 0 & 0 & 1 \\ 1 & 0 & 0 \\ 0 & 1 & 0 \end{pmatrix}\begin{pmatrix} 1 & 0 & 0 \\ 0 & \sin\alpha & \cos\alpha \\ 0 & -\cos\alpha & \sin\alpha \end{pmatrix} = \begin{pmatrix} 0 & -\cos\alpha & \sin\alpha \\ 1 & 0 & 0 \\ 0 & \sin\alpha & \cos\alpha \end{pmatrix}.$$

(4)

Note that since we started with $\alpha = 90°$, we have to adjust the right-hand rotation matrix to refer to a rotation of $90° - \alpha$. Figure 1 shows the geometry with $\alpha$ between 0 and 90° and we see that (4) gets the sign right.

The next step is to apply the in-plane $\gamma$ rotation which mixes $\hat{x}$ and $\hat{y}$:

$$\begin{pmatrix} \hat{x} \\ \hat{y} \\ \hat{z} \end{pmatrix}\bigg|_{\alpha,0,\gamma} = \begin{pmatrix} \cos\gamma & \sin\gamma & 0 \\ -\sin\gamma & \cos\gamma & 0 \\ 0 & 0 & 1 \end{pmatrix}\begin{pmatrix} \hat{x} \\ \hat{y} \\ \hat{z} \end{pmatrix}\bigg|_{90°,0,0}, \quad (5)$$

Using Mathematica to do the algebra results in the final matrix:

$$R_{\alpha 0\gamma} = \begin{pmatrix} \hat{x} \\ \hat{y} \\ \hat{z} \end{pmatrix}\bigg|_{\alpha,0,\gamma} = \begin{pmatrix} \sin\gamma & -\cos\alpha\cos\gamma & \sin\alpha\cos\gamma \\ \cos\gamma & \cos\alpha\sin\gamma & -\sin\alpha\sin\gamma \\ 0 & \sin\alpha & \cos\alpha \end{pmatrix}. \quad (6)$$

This is the matrix which, when multiplying a vector in lab coordinates, yields its equivalent in sample coordinates. Even if the assumptions about the order of rotations is wrong, the use of this rotation matrix should still let us determine whether a given set of measurements is enough to perform reconstructions. The inverse of this matrix is:

$$R_{\alpha 0\gamma}^{-1} = \begin{pmatrix} \sin\gamma & \cos\gamma & 0 \\ -\cos\alpha\cos\gamma & \sin\alpha\sin\gamma & \sin\alpha \\ \cos\gamma\sin\alpha & -\sin\alpha\sin\gamma & \cos\alpha \end{pmatrix} = R_{\alpha 0\gamma}^{T}. \quad (7)$$

Next, we will need the polarization and the beam direction, expressed in the sample frame of reference, in which $\hat{x}, \hat{y}, \hat{z}$ are the basis vectors. The beam direction, which I have been denoting by $\hat{n}$, is given by the third column of the rotation matrix:

$$\hat{n} \cdot \begin{pmatrix} \hat{x} \\ \hat{y} \\ \hat{z} \end{pmatrix} = \begin{pmatrix} \sin\alpha \cos\gamma \\ -\sin\alpha \sin\gamma \\ \cos\alpha \end{pmatrix} \quad (8)$$

and the polarization vector, again in sample coordinates, is

$$\hat{e} = R_{\alpha\gamma} \hat{E}_\chi = \begin{pmatrix} \sin\gamma \cos\chi - \cos\gamma \cos\alpha \sin\chi \\ \cos\gamma \cos\chi + \sin\gamma \cos\alpha \sin\chi \\ \sin\alpha \sin\chi \end{pmatrix}. \quad (9)$$

With the geometry in place, we can put in the physics of dichroism. What follows is a long string of formulas set out here for later use. For MCD, the effective absorption coefficient is linearly-related to the dot product of the beam direction and the magnetization, so can be expressed as[1]:

$$\mu(\vec{x},\gamma,s) = \mu_{nd}(\vec{x}) + s\vec{m}(x) \cdot \hat{n}(\gamma) \quad (10)$$

where $s = \pm 1$ depending on the handedness of the polarization, $\vec{m}$ is the product of the magnetization and a dichroism proportionality constant and $\mu_{nd}$ is a background, non-dichroic absorption. Just from imaging, it's impossible to separate the magnitude of the magnetization from the amount of dichroism a given magnetization produces. However, it's likely in many systems that this response coefficient is proportional to $\mu_{nd}$. In particular, if $\mu_{nd} = 0$ for a voxel, that voxel is empty so $\vec{m}$ must also vanish. If we represent $\vec{m}$ in polar coordinates, $\vec{m} = m(\sin\theta\cos\phi, \sin\theta\sin\phi, \cos\theta)$, we have

$$\mu = \mu_0 + sm(\sin\alpha\sin\theta\cos(\gamma+\phi) + \cos\alpha\cos\theta). \quad (11)$$

Note that if we subtract the projections done at one polarization from corresponding projections done at the other, we remove $\mu_{nd}$, leaving the dichroic contribution. Decomposing this into single trig functions, we have

$$\begin{aligned} \mu &= C_0 + C_1 \cos\gamma + S_1 \sin\gamma \\ C_0 &= \mu_{nd} + sm_z \cos\alpha \\ C_1 &= sm_x \sin\alpha \\ S_1 &= -sm_y \sin\alpha \end{aligned} \quad (12)$$

or, equivalently,

$$\begin{aligned} \mu &= \mu_{-1} e^{-i\gamma} + \mu_0 + \mu_1 e^{i\gamma} \\ \mu_{-1} &= s\sin\alpha(m_x + im_y)/2 \\ \mu_0 &= \mu_{nd} + sm_z \cos\alpha \\ \mu_1 &= s\sin\alpha(m_x - im_y)/2 \end{aligned} \quad (13)$$

Here, $m_{x,y,z}$ are the components of $\vec{m}$ in sample coordinates. We see that in CT mode ($\alpha = 90°$), we lose information about one of the components of the magnetization, $m_z$. This is because in CT mode, the sample $z$ axis is always perpendicular to the beam, so there is no dichroism due to magnetization in that direction. In CL mode, if we can get the three moments we can get the scaled magnetization because unless $\alpha$ is $0°$ or $90°$, the set of linear equations (13) is non-singular.

For linear dichroism, the formalism is more complex. Many materials of interest are optically uniaxial due to their crystallinity (calcite, for instance) or nearly so (aragonite and vaterite). Similarly, antiferromagnetism induces a linear dichroism on resonance. If we assume uniaxiality with an orientation unit vector $\hat{p}$, then the absorption may be described according to Malus' law as

$$\mu = \mu_{nd} + \mu_d (\hat{p} \cdot \hat{e})^2 \tag{14}$$

where $\mu_d$ measures the strength of the dichroism. Expressing the orientation in polar coordinates as

$$\hat{p} = \hat{x} \sin\theta \cos\phi + \hat{y} \sin\theta \sin\phi + \hat{z} \cos\theta \tag{15}$$

we have,

$$\mu = \mu_{nd} + \mu_d (\sin\chi \cos\theta + \cos\chi \sin\theta \sin(\gamma + \phi))^2 \;, \tag{16}$$

which looks similar to the equation we had for MCD, (11), except that the angle-dependent term is squared. The math simplifies a bit if we assume that $\mu_d$ has the same sign everywhere and re-express the dichroism as

$$\vec{p}' = \sqrt{|\mu_d|} \hat{p} \tag{17}$$

which makes the basic equation read

$$\mu = \mu_{nd} \pm (\vec{p}' \cdot \hat{e})^2 \tag{18}$$

where the $\pm$ sign is there to deal with the cases of positive or negative dichroism. In that case, we have

$$\mu = C_0 + C_1 \cos\gamma + C_2 \cos 2\gamma + S_1 \sin\gamma + S_2 \sin 2\gamma$$

$$C_0 = \mu_{nd} \pm (3(p_x'^2 + p_y'^2) + 2p_z'^2 + (p_x'^2 + p_y'^2 - 2p_z'^2)(\cos 2\alpha + \sin^2\alpha \cos 2\chi))/8$$

$$C_1 = \pm 2 p_z' \sin\alpha \sin\chi (p_y' \cos\chi - p_x' \sin\chi \cos\alpha)$$

$$S_1 = \pm 2 p_z' \sin\alpha \sin\chi (p_x' \cos\chi + p_y' \sin\chi \cos\alpha) \tag{19}$$

$$C_2 = \pm \frac{1}{8}(p_y'^2 - p_x'^2)\left((3+\cos 2\alpha)\cos 2\chi + 2\sin^2\alpha\right) \mp p_x' p_y' \cos\alpha \sin 2\chi$$

$$S_2 = \pm \left(\frac{1}{2}(p_y'^2 - p_x'^2)\cos\alpha \sin 2\chi + \frac{p_x' p_y'}{4}\left((3+\cos 2\alpha)\cos 2\chi + 2\sin^2\alpha\right)\right)$$

or in exponential notation,

$$\mu = \mu_{-2}e^{-2i\gamma} + \mu_{-1}e^{-i\gamma} + \mu_0 + \mu_1 e^{i\gamma} + \mu_2 e^{2i\gamma}$$

$$\mu_{-2} = \mp \frac{p'_x - ip'_y}{4}(\cos\chi + i\cos\alpha \sin\chi)^2$$

$$\mu_{-1} = \mp(p'_x - ip'_y)p'_z \sin\alpha \sin\chi(\cos\alpha \sin\chi - i\cos\chi)$$

$$\mu_0 = \mu_{nd} \pm (3(p'^2_x + p'^2_y) + 2p'^2_z + \quad\quad\quad (20)$$
$$(p'^2_x + p'^2_y - 2p'^2_z)(\cos 2\alpha + \sin^2\alpha \cos 2\chi))/8$$

$$\mu_1 = \mp(p'_x + ip'_y)p'_z \sin\alpha \sin\chi(\cos\alpha \sin\chi + i\cos\chi)$$

$$\mu_2 = \mp \frac{p'_x + ip'_y}{4}(\cos\chi - i\cos\alpha \sin\chi)^2$$

Note that with Mathematica, the easiest way to extract these coefficients is to evaluate weighted integrals such as $\mu_2 = (1/2\pi)\int_0^{2\pi} \mu(\gamma)\exp(-2i\gamma)$.

From the above, we see that from the ratio of, for instance, $\mu_2$ to $\mu_1$ we could extract $p'_z$, and from $\mu_{\pm 1}$, $p'_{x,y}$ and then from $\mu_0$, the non-dichroic background $\mu_{nd}$. This only works in the case of laminography ($\alpha \neq 90°$) and an oblique polarization. Thus, if we can somehow reconstruct the moments, we can get the information we want, in most cases without needing to switch polarizations (XLD) or tilt angles (MCD). We will see later on, however, that we can't actually get all the moments without adding extra information.

There is another way to describe XLD which is mathematically more elegant than the above, but does not easily enforce the assumption of uniaxiality. This description also covers the more general case of biaxial materials such as orthorhombic or monoclinic crystals. The absorption is described by a symmetric tensor, which is proportional to the imaginary part of the dielectric tensor:

$$\mu = \hat{e}^T \cdot Q \cdot \hat{e} = \hat{e}^T \cdot \begin{pmatrix} Q_1 & Q_4 & Q_5 \\ Q_4 & Q_2 & Q_6 \\ Q_5 & Q_6 & Q_3 \end{pmatrix} \cdot \hat{e} \quad\quad (21)$$

so the unknowns form a 6-component vector $(Q_1, Q_2, Q_3, Q_4, Q_5, Q_6)$. This one-index notation is what's used in other fields such as solid mechanics (strain, stress). For a uniaxial material (equivalently, for antiferromagetism), the tensor has two equal eigenvalues. The twin eigenvalue is the absorption coefficient perpendicular to the axis and the lone eigenvalue is the one parallel. For example, if the axis is along $\hat{x}$, then the 6-component vector is $(\mu_\perp, \mu_\perp, \mu_\parallel, 0, 0, 0)$. The isotropically-averaged absorption coefficient is

$\mu_{iso} = TrQ/3 = (Q_1 + Q_2 + Q_3)/3.$ The tensor, for the uniaxial case, can be expressed as

$$Q = \mu_{nd} + \mu_d \hat{p}^T \hat{p}. \quad\quad (22)$$

The advantage of this representation is that, unlike the uniaxial vector representation, the equations are linear in the unknowns so matrix algebra can be used, just as in the MCD case. Also, there's no $\pm$ to keep track of. The moments are thus,

$$\begin{pmatrix} \mu_{-2} \\ \mu_{-1} \\ \mu_0 \\ \mu_1 \\ \mu_2 \end{pmatrix}^T = S_{XLD}(\chi) \begin{pmatrix} Q_1 \\ Q_2 \\ Q_3 \\ Q_4 \\ Q_5 \\ Q_6 \end{pmatrix}$$

$$S_{XLD}(\chi) = \begin{pmatrix} \dfrac{-(c_\chi + is_\chi c_a)^2}{4} & \dfrac{(c_\chi + is_\chi c_a)^2}{4} & 0 & \dfrac{i(c_\chi + is_\chi c_a)^2}{2} & 0 & 0 \\ 0 & 0 & 0 & 0 & is_a s_\chi (c_\chi + ic_a s_\chi) & s_a s_\chi (c_\chi + ic_a s_\chi) \\ \dfrac{c_\chi^2 + c_a^2 s_\chi^2}{2} & \dfrac{c_\chi^2 + c_a^2 s_\chi^2}{2} & s_a^2 s_\chi^2 & 0 & 0 & 0 \\ 0 & 0 & 0 & 0 & -is_a s_\chi (c_\chi - ic_a s_\chi) & s_a s_\chi (c_\chi - ic_a s_\chi) \\ \dfrac{-(c_\chi - is_\chi c_a)^2}{4} & \dfrac{(c_\chi - is_\chi c_a)^2}{4} & 0 & \dfrac{-i(c_\chi - is_\chi c_a)^2}{2} & 0 & 0 \end{pmatrix}$$

(23)

where $c_\chi = \cos\chi$; $s_\chi = \sin\chi$; $c_\alpha = \cos\alpha$; $s_\alpha = \sin\alpha$. For an isotropic sample, $Q_{1-3}$ are equal and $Q_{4-6}$ vanish, which results in $\mu_m = 0$ for $m \neq 0$ and $\mu_0$ independent of $\chi, \alpha$, recovering the isotropy of $\mu$. In MCD we have

$$\begin{pmatrix} \Delta\mu_{-1} \\ \Delta\mu_0 \\ \Delta\mu_1 \end{pmatrix} = S_{MCD} \begin{pmatrix} m_x \\ m_y \\ m_z \end{pmatrix} = (1/2) \begin{pmatrix} \sin\alpha & -i\sin\alpha & 0 \\ 0 & 0 & 2\cos\alpha \\ \sin\alpha & i\sin\alpha & 0 \end{pmatrix} \begin{pmatrix} m_x \\ m_y \\ m_z \end{pmatrix} \quad (24)$$

where $\Delta\mu_m = \mu_m\big|_{s=+1} - \mu_m\big|_{s=-1}$ is the dichroic contrast. We will use these matrices later.

For the MCD case, if we can measure all three moments, we can get all components of the magnetization because we can solve for $\vec{m}$ in (24), provided $\alpha$ is not 0° or 90°. However, for XLD without some additional assumption or information such as perhaps another polarization, we can't because there are 6 unknowns and only 5 observables. One possibility is to constrain to uniaxiality. Unfortunately, this constraint is a non-linear condition, mainly that two of the eigenvalues of $Q$ are equal. Such a condition can't be implemented simply by, for instance, adding another row to $S$. Another possibility is to take another scan at an energy at which there is no dichroism, then assume based on the composition of the sample a scaling relation between the observed absorption coefficient at that energy and the isotropic part at the energy at which the dichroic CT/CL is performed. That effectively gives the trace of $Q$, adding a sixth component to the moments vector and a sixth row to $S$. We can test this by adding another row to $S$ consisting of (1,1,1,0,0,0) and asking Mathematica for the rank of the now 6x6 matrix. The result is 6, showing that adding the constraint on the trace of $Q$ fixes the under-determination problem, given knowledge of the moments. However, this only works if the sample is homogeneous and its optical properties are very well-known, perhaps from measurements on a polycrystal.

If we use another polarization, then we have 10 observables and 6 unknowns, so one might think that there are plenty of constraints. Actually, the problem is just determined. Proof: Set up the equations for two polarizations:

$$\begin{pmatrix} M(\chi_1) \\ M(\chi_2) \end{pmatrix} = \begin{pmatrix} S(\chi_1) \\ S(\chi_2) \end{pmatrix} \mathbf{Q} \qquad (25)$$

where $M(\chi)$ is the column vector of moments for polarization $\chi$ and $\mathbf{Q}$ is the column vector of absorption-tensor values $Q_{1...6}$. The matrix that multiplies $\mathbf{Q}$ has 10 rows and 6 columns and can be assembled in Mathematica with the Catenate[ ] function. The MatrixRank[ ] function then reports that the rank is 6, just enough to solve for $\mathbf{Q}$.

At this point, it would appear that we can find all three components of the scaled magnetization (CL mode only) or all six components of the absorption tensor for XLD using two polarization (LCP/RCP for MCD, two angles for XLD). However, this is contingent on being able to measure all three (MCD) or five (XLD) moments. We will see that this is actually not possible; the set of projections does not yield enough information. This deficit is over and above the missing cone in Fourier space that's inevitable in CL and which we will discuss later.

## 4. Back-projections and filtered back-projections

So far, it is now established that for MCD, reconstructing three moments of the absorption will provide all the information we need. For XLD, five moments would be sufficient if another constraint, such as uniaxiality, could be imposed. Now we need to figure out how to do this reconstruction, and whether and under what circumstances it's possible. We have already been using assumption A1 (ability to describe transmission in terms of a local absorption) and A5 (collimated beam). From here on, we will use A3 (no missing angles) and A4 (back-projections map into projections). However, before getting into the details of reconstruction, I will digress a bit to discuss the missing information in laminography, even for isotropic materials.

### 4.1 Missing information in laminography

One way to see that there must be missing information in laminography is to consider the method applied to a sample which is layered normal to the rotation axis but is uniform in-plane. For this case, rotation of the sample produces no variation of absorption; all projections are the same. Therefore, the layering is completely unresolved and all one gets is the total absorption through the sample.

The effect may be described by considering the back-projection in Fourier space. Recall that the back-projection is the sum of many single back-projections. Each of these has the form of a bundle of lines of density extended along the beam direction for the given projection. The Fourier transform of such a bundle is a sheet oriented perpendicular to the beam direction. When the sample is rotated, the sheet, in the sample's reference frame, rotates about the sample rotation axis, resulting in a double cone within which there is no intensity. Thus, Fourier components within this cone are not sampled and no information is available about them. The mathematics of this phenomenon are shown explicitly in [17] and [18]. Examples of the reconstruction artefacts produced by this missing cone are shown in [19] in their demonstration of cone-beam laminography on a laboratory apparatus. Note that although the missing cone was demonstrated mathematically in the context of filtered back-projection reconstruction, the effect is independent of reconstruction method because it's a real gap in the available information. Theorem A4 shows that no reconstruction algorithm will dig up information that filtered back-projection can't. As an example, the reconstructions shown in [19] were done using an iterative least-squares algorithm, not filtered back-projection, yet show missing-cone artefacts. However, with prior information about the unknown sample, it is possible to mitigate these effects. For example, it is shown in [20] that knowledge of the minimum and maximum values of the refractive index (they reconstructed phase contrast) can be used with an iterative modification of FBP to reduce the missing-cone artefacts.

*4.2 Reconstruction of moments*

We have seen that the unknown object may be described in terms of effective absorption coefficients which oscillate as a function of rotation angle $\gamma$. Suppose we want to derive a generalization of filtered back-projection which can reconstruct not just an angle-independent $\mu$ but all the moments $\{\mu_i\}$. The back-projection corresponding to a single point of isotropic absorption at position $\vec{x} = \vec{x}_0$ looks like a double cone of non-zero values with the vertex at position $\vec{x}_0$, oriented along the rotation axis. The intensity is concentrated at this vertex because the lines from the different angles crowd together, thus producing a power-law singularity. This is why the unfiltered back-projection resembles the object but with a 'haze' around each point. Now consider the (unphysical) case in which only one of the non-zero moments appears, $\mu_i(\vec{x}) = \delta(\vec{x} - \vec{x}_0)\delta_{im}$. The back-projection produced by the "object" is still a double cone, but modulated with angle by a factor of $e^{im\gamma}$. Unlike the back-projection from an isotropic point absorber, the double cone with $m \neq 0$ does not have a concentration at the vertex. The appearance of the back-projections of point objects was verified by simulation using the ASTRA toolbox [21]. If we perform the back-projection of an object having intensity in all the moments $\mu_m, m \in [-2, 2]$ (XLD) or $m \in [-1, 1]$ (MCD), we get a complex mixture of contributions from all the $\mu_i$. The $\mu_0$ component will still be concentrated at points where $\mu_0$ is high, but simulations show that the other moments contribute significantly.

Now, if we perform the back-projection by first weighting each projection by $e^{-im\gamma}$, we change the $\mu_m$ moment into what looks like $\mu_0$, somewhat the way a lock-in amplifier converts an oscillating signal into a DC one by mixing it with a reference of the same frequency. Thus, we could imagine performing a series of filtered back-projections on weighted projections, then solving for the moments. This process is illustrated graphically in Figure 2. In this example, real-valued trig-function weightings are shown, rather than complex exponentials.

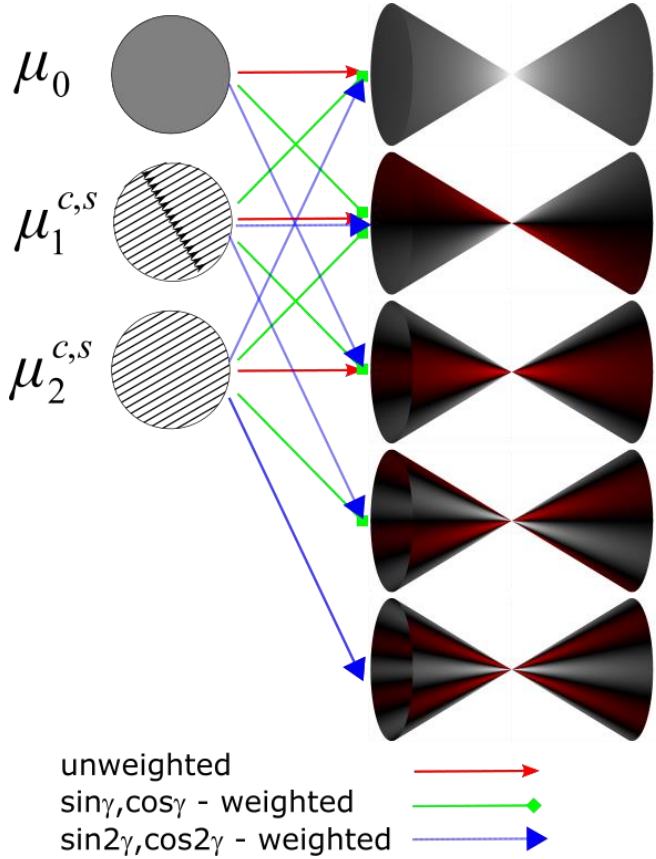

Fig. 2. A graphical representation of the effect of performing $\sin, \cos(n\gamma)$-weighting on the contributions to weighted back-projections from the components of the absorption density. The double cones represent weighted projections of a point object with positive values shown as grey (brighter = higher), negative values shown as red and 0 as black. Note that the resulting angular components in the projections go up to $n=4$. The weightings are shown as arrows connecting moments of the absorption (the circles on the left) with the resulting contributions to the back-projections.

Here, the uniformly-shaded disk represents isotropic absorption, the single-headed arrows represent moments of order 1 and the lines with no arrows represent moments of order 2. The double cones represent back-projections of point objects with varying angular dependence. We see that the contributions of moments of order up to $\pm 2$, when weighted, yield contributions to order $\pm 4$. Thus, in order to go from observables (the back-projections) to unknowns (the moments), we must perform what is effectively a matrix inversion.

### 4.2.1 Generalized filtered back-projection

In ordinary filtered back-projection, the filtering is done by computing the FT of the double-conical back-projection and using the inverse (where it's not 0) as a deconvolution filter. Here, we generalize to a multicomponent problem. For this purpose, it is easier to work with complex components, as in Eq. (20), than sines and cosines as in Figure 2. The weighted projections from an isotropic point object at $\vec{x} = 0$ are given by:

$$P_n(\vec{x}) = \int_0^{2\pi} d\gamma \exp(in\gamma) \int \delta(\vec{x} - \hat{n}(\gamma)l) dl \qquad (26)$$

where $\hat{n}(\gamma)$ is the direction along the ray when the projection angle is $\gamma$ and $l$ is a path length along the ray. The projections from an extended object are sums of convolutions of $P_n$ with the angular components $\mu_m$, $m = -2...2$:

$$p_n(\vec{x}) = \sum_{m=-2}^{2} \int_0^{2\pi} d\gamma \exp(i(n+m)\gamma) \int dl \mu_m(\vec{x} + l\hat{n}(\gamma))$$
$$= \sum_{m=-2}^{2} (\mu_m \otimes P_{n+m}) \quad (27)$$

where $P_n$ is given by

$$P_n(x) = \int_0^{2\pi} d\gamma e^{in\gamma} \int_{-\infty}^{\infty} dl \delta(\vec{x} - \hat{n}(\gamma)l) \quad . \quad (28)$$

What we need is the FT of this, in order to do the deconvolution. The back-projection is a sum of lines all going through the origin. Therefore, the FT is a sum of sheets in $k$-space, produced by taking one sheet, tilted with its normal tilted at an angle $\alpha$ to the rotation ($\hat{z}$) axis, and weighted by the $e^{im\gamma}$ factor. Thus, we can write the FT of weighted back-projection $P_m$ as

$$F_m = \frac{1}{2\pi} \int_0^{2\pi} d\gamma e^{im\gamma} \delta(k_z \cos\alpha + \sin\alpha[k_x \cos\gamma + k_y \sin\gamma])$$
$$= \frac{1}{2\pi} \int_0^{2\pi} d\gamma e^{im\gamma} \delta(k_z \cos\alpha + \sin k_t \cos(\gamma - \psi)) \quad (29)$$
$$= \frac{e^{im\psi}}{2\pi} \int_0^{2\pi} d\gamma e^{im\gamma} \delta(k_z \cos\alpha + \sin\alpha k_t \cos\gamma)$$

with the substitution $k_t = \sqrt{k_x^2 + k_y^2}$ and $\psi$ is the azimuthal angle of $\vec{k}$, thus $k_x = k_t \cos\psi$; $k_y = k_t \sin\psi$. Next, we express the delta function as the well-known integral $\delta(x) = (1/2\pi) \int_{-\infty}^{\infty} dw \exp(ixw)$ and reverse the order of integrations:

$$F_m = \frac{e^{im\psi}}{(2\pi)^2} \int_{-\infty}^{\infty} dw \exp(iwk_z \cos\alpha) \int_0^{2\pi} d\gamma \exp(i(m\gamma + wk_t \sin\alpha \cos\gamma)) .$$
$$(30)$$

The second integral is one of the many integral forms of the $J$ Bessel function, so:

$$F_m = \frac{e^{im\psi}}{2\pi i^m} \int_{-\infty}^{\infty} dw \exp(iwk_z \cos\alpha) J_m(wk_t \sin\alpha) \quad (31)$$

The form for $m=0$ is the one presented by[18]. From there, $F_m$ for non-0 $m$ can be evaluated by integration by parts and the use of a recursion relation:

$$F_m(\vec{k}) = \begin{cases} \dfrac{e^{im\psi}}{\pi} \dfrac{T_{|m|}(z/t)}{\sqrt{t^2 - z^2}} & t \geq |z| \\ 0 & t < |z| \end{cases} \tag{32}$$

where $z = k_z \cos\alpha$; $t = k_t \sin\alpha$, $T_m(x)$ is the Chebyshev polynomial of the first kind, and this form is true for negative $m$ as well as positive. The missing cone falls naturally out of the math without needing to be put in by hand. The result for $m=0$ is the one well-known for filtered back-projection, derived for laminography in [18].

We need to be careful in evaluating $F$ for $k_z = 0$. At $k_t = k_z = 0$, the function is undefined because $\lim_{k_t \to 0} \lim_{k_z \to 0} F_m = 0$ but $\lim_{k_z \to 0} \lim_{k_t \to 0} F_m$ is infinite. It seems to be standard to take the value of the reciprocal transform ($F_m^{-1}$ in our case) to be 0 at the origin of $k$-space. For $k_z = 0, k_t \neq 0$ the integral is:

$$F_m(k_t, k_z = 0) = \begin{cases} \dfrac{e^{im\psi}(-1)^{m/2}}{\pi k_t \sin\alpha} & m \text{ even} \\ 0 & m \text{ odd} \end{cases}. \tag{33}$$

Similarly, in CT mode, $z$ is 0 everywhere, so $F_m$ is also given by (33). Eq. 4Now that we have the FTs. we find that the FT of the weighted back-projections is:

$$\overline{p}_n(\vec{k}) = \sum_{m=-2}^{2} F_{n+m}(\vec{k}) \overline{\mu}_m \tag{34}$$

where the over-bar represents Fourier transformation. This is a matrix equation in thin disguise, with the unknown vector being $\{\overline{\mu}_m\}$. Thus, the procedure would be to do 3D FFTs of the set of weighted back-projections, then, for all points in $k$-space not in the missing cone, multiply the vector of FFTs by the inverse matrix, leaving the missing-cone values set to 0, then inverting the FFT. The matrix algebra becomes easier if we re-express (34) in a form that separates out the phase factors:

$$\overline{p}_n(\vec{k}) = \sum_{m=-2}^{2} e^{in\psi} G_{n+m}(\vec{k}) e^{im\psi} \overline{\mu}_m \tag{35}$$

where $G_n = e^{-in\gamma} F_n$; $G_{-n} = G_n$. Thus, if we multiply $\overline{p}_n$ by $e^{-in\psi}$ and $\overline{\mu}_n$ by $e^{in\psi}$, we get a form with a real, symmetric matrix:

$$\tilde{P} \equiv \begin{pmatrix} \tilde{p}_{-2} \\ \tilde{p}_{-1} \\ \tilde{p}_0 \\ \tilde{p}_1 \\ \tilde{p}_2 \end{pmatrix}^T = \begin{pmatrix} G_0 & G_1 & G_2 & G_3 & G_4 \\ G_1 & G_0 & G_1 & G_2 & G_3 \\ G_2 & G_1 & G_0 & G_1 & G_2 \\ G_3 & G_2 & G_1 & G_0 & G_1 \\ G_4 & G_3 & G_2 & G_1 & G_0 \end{pmatrix} \begin{pmatrix} \tilde{\mu}_{-2} \\ \tilde{\mu}_{-1} \\ \tilde{\mu}_0 \\ \tilde{\mu}_1 \\ \tilde{\mu}_2 \end{pmatrix} \equiv \tilde{G}\tilde{M} \tag{36}$$

where quantities with a tilde are the $e^{\pm im\psi}$-multiplied over-barred quantities and the capital-tilde'd variables $\tilde{P}, \tilde{M}$ are the phase-multiplied variables $\tilde{p}_i, \tilde{\mu}_i$, piled vertically as in (36).

There is a special case which needs to be considered, which is that of $\vec{k} = 0$, which will be one of the points sampled in the discrete FT. The value of the FFT at the 0 point is the sum over the imaged volume of the operand. Consider a 'sample' which is uniform within some volume outside which it's empty, and has only an $m$'th moment, which means that it has an unphysical $\mu = \exp(im\gamma)$. Then, for the LHS of (36) we have $\tilde{\mu}_n = \delta_{nm}$ and $\tilde{p}_n = C\delta_{nm}$ where $C$ is a constant. To evaluate this constant, it's probably easiest to simulate the back-projections through a uniformly-filled volume and sum up all voxels of the back-projection. Since the proportionality of $\tilde{\mu}_n$ and $\tilde{p}_n$ holds for all values, it follows that the matrix in (36) must reduce to the identity matrix times $C$ for the $\vec{k} = 0$ point.

All of the above is aimed at the problem of reconstructing the moments $\mu_m$ of the absorption coefficients. From that, we can go back to the formalism starting at (10) to get the material properties represented by $\vec{m}$ or $Q_i$. For instance, the development leading to (20) relates the moments to a solution for uniaxial linear dichroism with a single oblique polarization in laminography, while that leading to (13) shows that the scaled magnetization can be solved for with two polarizations in CL mode.

Summarizing, we find that if we can reconstruct the moments $\mu_m$ of the absorption coefficients, for instance by a weighted back-projection method, we could reconstruct the orientation and magnitude of the dichroism. With what set of measurements can we pull out all the moments uniquely?

### 4.2.2 Missing information

There is a major problem with the above approach if one tries to use it with a single polarization and tilt. It turns out that the $\tilde{G}$ matrix in (36) is singular. That means that it's possible to add to the $\tilde{\mu}_m$ coefficients an extra contribution $A(\vec{k})u_m(z,t)$ where $A_m(\vec{k})$ is any function in $k$-space and $u_m(z,t)$ is one of the eigenvectors corresponding to a 0 eigenvalue (there is one for the 3x3 MCD case and three for XLD) of the matrix to be inverted. Several speculative workarounds can be proposed. One possibility is to solve the equation in the singular-value decomposition (SVD) sense, which gives *a* solution, and one with the smallest norm, but probably not *the* solution. It seems from numerical experiments that 'extra' components yield 'bleed' outside the boundaries of the object. Such 'bleed' may be expected to increase the norm of the solution and so not appear in the SVD solution. Yes, this is a weak argument. Another possible solution is to ask that outside of some support volume, all components of $\mu$ are zero. How would we get that support volume? Perhaps one could use a conventional method of reconstruction other than filtered back-projection, maybe at an energy at which dichroism is negligible. Some version of shrink-wrapping might also be possible. Thus, we can frame the problem as a minimization problem, with the norm of the absorption tensor outside the support volume as an additional cost function.

For XLD, one extra bit of information we have is that for many, if not most cases of interest, the optical properties are uniaxial, so the absorption coefficient tensor is given by (22). Counting up degrees of freedom yields four independent variables, i.e. two constraints. Since the matrix for XLD has three vanishing eigenvalues, it appears that uniaxiality does not provide enough constraint to remove the degeneracy. Also, the relationship between the weighted moments, in real space, is not simply linear, so linear methods don't work, as shown in (19).

Suppose we add polarizations to get more information in XLD. As we did previously in discussing the connection between material parameters and moments of $\mu$, we can make a pile of the matrices connecting $\{\tilde{p}_m\}$ and $\mathbf{Q}$ and test for rank. The resulting equation, in block-matrix form, is

$$\begin{pmatrix} \tilde{P}(\chi_1) \\ \ldots \\ \tilde{P}(\chi_{n_\chi}) \end{pmatrix} = \begin{pmatrix} \tilde{G}\tilde{S}(\chi_1) \\ \ldots \\ \tilde{G}\tilde{S}(\chi_{n_\chi}) \end{pmatrix} (\overline{Q}) \tag{37}$$

where $\overline{Q}$ is the set of 6 independent tensor components of the absorption in Fourier space, $n_\chi$ is the number of polarizations and tilts and $\tilde{S}(\chi)$ is the $k$-independent 5x6 matrix (from (23)) which relates the tensor components to the angular components, multiplied on the left by $e^{im\gamma}$. Does that solve the problem? Again, we can ask Mathematica for the rank of the piled matrix. We find that if all the measurements are done at the same tilt, the rank is 2 with one polarization, 4 with two and 5 for three or more. To get to a rank of 6, we need to have at least one of the measurement done with a different tilt than the others. For a single polarization, two tilts yields a rank of 4 and three gets to 6. Thus, we need at least three measurements, either with three different polarizations and at least two distinct tilt values, or three distinct tilts at a single polarization.

It should be noted that since the matrix has more rows than columns, we're now in the domain of a least-squares solution, though one of manageable size for each $k$-point, rather than the global one with hundreds of trillions of elements.

For MCD, we do have a missing-information problem as was noted by [22], who showed that CL or CT done at a single orientation does not yield a unique solution for the magnetization. That result is reproduced in this work. While it's true that the rank of $S_{MCD}$ is 3 for CL, the rank of the 3x3 matrix corresponding to the $G$-matrix in (36) is 2, so we can't solve for all components of $\vec{m}$. We can't add polarizations because in MCD we only have two, and we're using both to get rid of the non-dichroic term. This rank is preserved on multiplying on the right by $S_{MCD}$. Thus, we need a constraint. On certain length scales, many materials have magnetizations either only in the plane of a film or perpendicular to it. If the film plane is the rotation plane, then the perpendicular case corresponds to $\vec{m} = (0, 0, m_z)$. In this case, tomography as we have been discussing it fails to deliver 3D information. If a CT geometry is used, the magnetization is perpendicular to the beam direction, so no dichroic signal will be observed. However, this case is somewhat artificial; the domain walls or other boundaries between areas of differing $m_z$ have local moments out of the normal direction, so the only information we can get comes from places where the constraint is violated. For the in-plane case, the nullspace of the $\tilde{G}S$ matrix contains one vector, which has components along $\hat{x}, \hat{z}$ directions. This implies that anything one might add to $\vec{m}(\vec{x})$ that still satisfies the linear equation of the reconstruction will inevitably have a component in the $\hat{z}$-direction, which we're not allowing. Unfortunately, many of the interesting structures we would want to look at such as skyrmions or hopfions obey no such constraints, as shown for instance in Figure 3 of [22].

One way around the problem is to use multiple tilt angles, by analogy with the multiple polarizations and tilts used for XLD. The calculation must be done carefully because the matrix $\tilde{G}$ depends on the tilt angle for a given point in Fourier space so each block in the pile has its own $\tilde{G}$. On doing the same kind of calculation for MCD as above, but with two different values of $\alpha$, we find a matrix rank of 3, meaning that two tilts are necessary and sufficient. One group [23, 24] varied both $\gamma$ and $\alpha$, but they inverted the role of the two angles, using only two $\gamma$ angles but scanning $\alpha$. Another group [22] performed magnetic tomography using a similar method to what's discussed here, with two tilt angles and 180 rotations. For XLD, doing the same calculation for a single polarization shows that we need three tilts. The singularity is removed with three tilts only if the polarization is neither horizontal nor vertical.

## 5. Conclusions, demonstration and discussion

I have laid out a framework for reconstructing laminographic images of anisotropic samples under conditions of variable polarizations, as a generalization of filtered back-projection. Using this, it was possible to determine that for XLD, tomograms at at least three polarizations are required for full information to be retrieved. Further, the tilt angles for all the polarizations must not be equal. Alternatively, three tomograms at different tilt angles α measured at a common polarization other than horizontal or vertical provide enough information. This result is relevant for the use of dichroic tomography on beamlines whose sources are bending magnets or horizontally-polarized undulators, both of which have polarization fixed in the horizontal plane relative to gravity. Thus, to perform multi-tilt XLD tomography, the α axis must be tilted at an angle to the horizontal, i.e. β must be non-zero. For MCD, it is necessary to image at at least two tilt angles. These conditions hold even if algorithms other than filtered back-projection are used, because the back-projections hold all the information available.

We can summarize the above math by assembling a weighted FBP algorithm by analogy with the standard scalar algorithm. The simplest scalar back-projection algorithm may be summarized as:

1. Create the back-projection from the sum of the individual back-projections, to get the back-projection.

2. Take the 3D Fourier transform

3. Divide this by the function $F_0(k)$ except where $F_0(k)$ vanishes. Zero out the FT where $F_0(k)$ vanishes.

4. Take the inverse Fourier transform. The result is an estimate for the desired distribution $\mu(\vec{x})$ subject to the distortions induced by the missing cone in Fourier space.

This method has a number of drawbacks, including the inefficient use of a 3D FT-inverse FT pair, but it serves to illustrate the point. The corresponding algorithm sketch for a tensor order parameter is:

1. For each polarization and tilt, create the weighted back-projections $\{\bar{p}_m\}$ from the sum of the back-projections for all angles.

2. Take the 3D Fourier transforms of the set of weighted back-projections. This yields
$$\begin{pmatrix} \bar{P}(\chi_1) \\ \cdots \\ \bar{P}(\chi_{n_\chi}) \end{pmatrix}, \text{ where } \chi \text{ refers to polarization and tilt.}$$

3. Solve the linear equation (37) for $k$-values outside the missing cone, yielding the Fourier transforms of the absorption tensor. Zero out the values inside the missing cone.

4. Take the inverse Fourier transforms to get the absorption tensor.

Another way to derive the matrix that goes into (37) is by direct simulation: place a point (1-voxel) object into the volume to be reconstructed and simulate the projections, then weighted back-projections and take the Fourier transforms. Do this for a set of 6 objects, each of which has one non-zero element of $Q$ and repeat for all conditions of polarization and tilt, yielding the set of $\{\bar{P}(\chi_i)(\vec{k})\}$ for $i \in [1, n_\chi]$. This method can be used when the sample geometry is not exactly as in Figure 1, for instance with different rotation axes, and also could also be used in setting up the forward model for real-space iterative methods.

In order to make the above more concrete and demonstrate that it works, an algorithm was implemented in Python using the ASTRA toolbox to simulate the projections from a known phantom and perform the back-projections. The program which did the reconstructions was separate from the one which did the reconstruction so can't "know" what the "right answer" is. The object chosen is a cube of a uniaxial material whose orientation varies in space in a helical fashion, like a cholesteric liquid crystal whose pitch axis is along a body diagonal of the cube and with a pitch equal to half the length of the diagonal. The absorption coefficient is 2 (arbitrary units) along the local optic axis and 1 perpendicular to it. The absorption tensor is given by (22) with $\mu_{nd} = \mu_d = 1$ inside the cube vanishes outside. The optic axis $\hat{p}$ is perpendicular to the body diagonal and rotates around it. The geometry and results are shown in Figure 3.

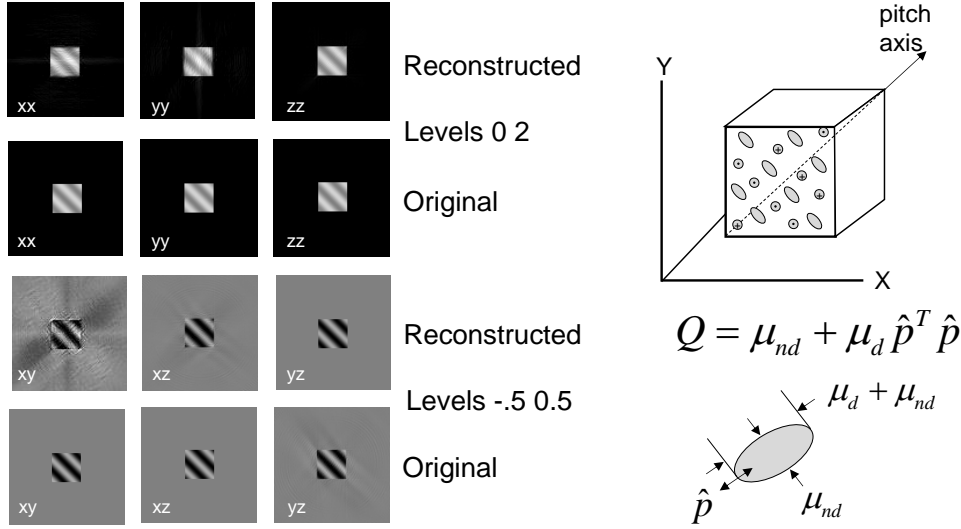

Fig, 3 A demonstration of a filtered weighted back-projection algorithm. The structure of the phantom is shown schematically on the right, with the ellipses representing the polarization dependence of the local absorption coefficient. The absorption tensor is given by (22) (shown in the Figure). The volume is 25x25x60 (XxYxZ) voxels across and is embedded in a 200x200x200 voxel volume, which is cropped to half-size in this Figure. The individual tensor components of the original and reconstructed phantoms are shown in grayscale, with consistent grayscales. In all images, the slice shown is at the midpoint of the cube ($z$ slice 100 of 200). The four conditions used were

$$\{(\chi, \alpha)\} = \{(0, 60°), (45°, 60°), (90°, 80°), (126°, 80°)\} .$$

The units for absorption coefficient are arbitrary, which is legitimate because the simulated imaging and reconstruction processes are both linear. The polarizations and tilts (listed in figure caption) were not chosen with any specific criterion in mind. We see that a reasonably-accurate reconstruction is obtained.

This simple algorithm is presented only to make clear how the preceding development relates to reconstruction. In actual practice, one would probably use something like SIRT and apply constraint methods such as were discussed above and not get involved with Fourier filtering at all. The reason for building up the formalism for the weighted FBP method is to examine the information content in a given set of projections, rather than to create a practical reconstruction algorithm.

In laminography, there is a loss of information in a cone in Fourier space which is, again, independent of the reconstruction method used. It is possible that constraints such as shrink-wrapping, uniaxiality (XLD) or non-negativity could help with this problem. Another possible approach is to use machine-learning techniques to constrain reconstructions to what's plausible, based on a training set of simulated tomograms constructed to have features similar to those found in real sample. This method was used to improve signal-to-noise and reduce required dosage in tomography by [25]. One may speculate that similar methods could be used to aid reconstruction in the case where an insufficient set of polarizations or tilts had been used for full recovery of the orientation. The training set must be carefully chosen so as not to impose constraints that the real sample doesn't obey.

For XLD, all of these results are contingent on the assumption that the absorption through a line in the sample may be considered to be the line integral of the local absorption along the line. The local absorption must be independent of the absorption anywhere else, a condition which requires that the polarization must be preserved as the beam propagates through the sample, something that does not happen if dichroism is strong. This problem does not occur in MCD because polarization is preserved during propagation.

**Acknowledgments.** The author acknowledges useful discussion and encouragement from D. A. Shapiro and D. Parkinson.

**Disclosures** The author declares no conflicts of interest.

**Data availability** Data files generated in the course of performing the simulation shown in Fig. 4. are not publically available but may be obtained from the author upon reasonable request.

**Supplemental document.** See Supplement 1 for supporting content.

**References**

1. P. Fischer, "X-ray imaging of magnetic structures," IEEE Transactions on Magnetics **51**, 1-31 (2014).
2. J. Stöhr and H. C. Siegmann, "Magnetism," Solid-State Sciences. Springer, Berlin, Heidelberg **5**(2006).
3. Y.-S. Yu, R. Celestre, B. Enders, K. Nowrouzi, H. Padmore, T. Warwick, J.-R. Jeong, and D. A. Shapiro, "Nanoscale Visualization of Magnetic Contrasts with Soft X-ray Spectro-Ptychography at the Advanced Light Source," Microscopy and Microanalysis **24**, 530-531 (2018).
4. K. G. Rana, R. L. Seeger, S. Ruiz-Gómez, R. Juge, Q. Zhang, V. T. Pham, M. Belmeguenai, S. Auffret, M. Foerster, and L. Aballe, "Room temperature skyrmions in an exchange biased antiferromagnet," arXiv preprint arXiv:2009.14796 (2020).
5. C. Donnelly, M. Guizar-Sicairos, V. Scagnoli, S. Gliga, M. Holler, J. Raabe, and L. J. Heyderman, "Three-dimensional magnetization structures revealed with X-ray vector nanotomography," Nature **547**, 328-331 (2017).
6. H. Braun and A. Hauck, "Tomographic reconstruction of vector fields," IEEE Transactions on signal processing **39**, 464-471 (1991).
7. C. A. Stifler, J. E. Jakes, J. D. North, D. R. Green, J. C. Weaver, and P. U. Gilbert, "Crystal misorientation correlates with hardness in tooth enamels," Acta Biomaterialia **120**, 124-134 (2021).
8. R. T. DeVol, R. A. Metzler, L. Kabalah-Amitai, B. Pokroy, Y. Politi, A. Gal, L. Addadi, S. Weiner, A. Fernandez-Martinez, R. Demichelis, J. D. Gale, J. Ihli, F. C. Meldrum, A. Z. Blonsky, C. E. Killian, C. B. Salling, A. T. Young, M. A. Marcus, A. Scholl, A. Doran, C. Jenkins, H. A. Bechtel, and P. U. P. A. Gilbert, "Oxygen spectroscopy and Polarization-dependent Imaging Contrast (PIC)-mapping of calcium carbonate minerals and biominerals," J. Phys. Chem. B **118**, 8449-8457 (2014).
9. H. Chang, M. A. Marcus, and S. Marchesini, "Analyzer-free linear dichroic ptychography," Journal of Applied Crystallography **53**(2020).
10. Y. H. Lo, J. Zhou, A. Rana, D. Morrill, C. Gentry, B. Enders, Y.-S. Yu, C.-Y. Sun, D. A. Shapiro, and R. W. Falcone, "X-ray linear dichroic ptychography," Proceedings of the National Academy of Sciences **118**(2021).
11. Z. Gao, M. Holler, M. Odstrcil, A. Menzel, M. Guizar-Sicairos, and J. Ihli, "Nanoscale crystal grain characterization via linear polarization X-ray ptychography," Chemical Communications **56**, 13373-13376 (2020).
12. J. Lüning, F. Nolting, A. Scholl, H. Ohldag, J. W. Seo, J. Fompeyrine, J.-P. Locquet, and J. Stöhr, "Determination of the antiferromagnetic spin axis in epitaxial LaFeO 3 films by x-ray magnetic linear dichroism spectroscopy," Physical Review B **67**, 214433 (2003).
13. D. Zhao, H. Höchst, and D. L. Huber, "Dielectric tensor formulation of magnetic dichroism sum rules," Journal of applied physics **84**, 2858-2860 (1998).
14. F. Pfeiffer, "X-ray ptychography," Nature Photonics **12**, 9-17 (2018).
15. A. van der Sluis and H. A. van der Vorst, "Numerical solution of large, sparse linear algebraic systems arising from tomographic problems," in *Seismic Tomography: With Applications in Global Seismology and Exploration Geophysics*, G. Nolet, ed. (Springer Netherlands, Dordrecht, 1987), pp. 49-83.
16. I. Wolfram Research, "Mathematica" (2021), retrieved https://www.wolfram.com/mathematica.
17. A. Myagotin, A. Voropaev, L. Helfen, D. Hänschke, and T. Baumbach, "Efficient volume reconstruction for parallel-beam computed laminography by filtered backprojection on multi-core clusters," IEEE Transactions on Image Processing **22**, 5348-5361 (2013).
18. H. Matsuo, A. Iwata, I. Horiba, and N. Suzumura, "Three-dimensional image reconstruction by digital tomo-synthesis using inverse filtering," IEEE transactions on medical imaging **12**, 307-313 (1993).

19. S. Fisher, D. Holmes, J. Jørgensen, P. Gajjar, J. Behnsen, W. Lionheart, and P. Withers, "Laminography in the lab: imaging planar objects using a conventional x-ray CT scanner," Measurement Science and Technology **30**, 035401 (2019).
20. S. Harasse, W. Yashiro, and A. Momose, "Iterative reconstruction in x-ray computed laminography from differential phase measurements," Optics Express **19**, 16560-16573 (2011).
21. W. Van Aarle, W. J. Palenstijn, J. Cant, E. Janssens, F. Bleichrodt, A. Dabravolski, J. De Beenhouwer, K. J. Batenburg, and J. Sijbers, "Fast and flexible X-ray tomography using the ASTRA toolbox," Optics express **24**, 25129-25147 (2016).
22. C. Donnelly, S. Gliga, V. Scagnoli, M. Holler, J. Raabe, L. J. Heyderman, and M. Guizar-Sicairos, "Tomographic reconstruction of a three-dimensional magnetization vector field," New Journal of Physics **20**, 083009 (2018).
23. A. Hierro-Rodriguez, C. Quirós, A. Sorrentino, L. M. Álvarez-Prado, J. Martín, J. M. Alameda, S. McVitie, E. Pereiro, M. Velez, and S. Ferrer, "Revealing 3D magnetization of thin films with soft X-ray tomography: magnetic singularities and topological charges," Nature communications **11**, 1-8 (2020).
24. A. Hierro-Rodriguez, D. Gürsoy, C. Phatak, C. Quirós, A. Sorrentino, L. M. Álvarez-Prado, M. Vélez, J. I. Martín, J. M. Alameda, and E. Pereiro, "3D reconstruction of magnetization from dichroic soft X-ray transmission tomography," Journal of synchrotron radiation **25**, 1144-1152 (2018).
25. D. M. Pelt, K. J. Batenburg, and J. A. Sethian, "Improving tomographic reconrtruction from limited data using mixed-scale dense convolutional neural networks," Journal of Imaging **4**, 128 (2018).

## supplemental information

### 6. Math notes

This section supplies some derivations not included in main text.

The first of these is how to derive Eq. 33 in the main text, the one which reads:

$$F_m(\vec{k}) = \begin{cases} \dfrac{e^{im\psi}}{\pi} \dfrac{T_{|m|}(z/t)}{\sqrt{t^2 - z^2}} & t \geq |z| \\ 0 & t < |z| \end{cases} \quad (38)$$

This starts with the definition:

$$F_m = \frac{e^{im\psi}}{2\pi i^m} \int_{-\infty}^{\infty} dw \exp(iwz) J_m(wt) \quad (39)$$

Mathematica has trouble with evaluating this integral because it ignores the possibility of having $t \geq |z|$ and so only yields 0 with a conditional. First, let us specialize to positive $m$. This is easy enough given that $J_{-n}(x) = (-1)^n J_n(x)$. Next, we consider the form for $m=0$, which is given by [18] and agrees with (38) since $T_0(x) = 0$. Next, we can evaluate $F_1$ using the relation [26] $J_1(x) = -J_0'(x)$. This yields

$$F_1 = \frac{e^{i\psi}}{2\pi i} \int_{-\infty}^{\infty} dw\, e^{iwz} J_1(wt) = -\frac{e^{i\psi}}{2\pi i} \int_{-\infty}^{\infty} dw\, e^{iwz} J_0'(wt)$$
$$= -\frac{e^{i\psi}}{2\pi i} \int_{-\infty}^{\infty} dw\, e^{iwz} \frac{1}{t} \frac{d}{dw} J_0(wt) \quad (40)$$

which can be integrated by parts to yield

$$F_1 = \frac{e^{i\psi}}{2\pi} \frac{z}{t} \int_{-\infty}^{\infty} dw\, e^{iwz} J_0(wt) = e^{i\psi}(z/t) F_0 \quad (41)$$

which again agrees with (38). This suggests that, aside from the phase factor, $F_m$ may be expressed as $F_0$ times a polynomial in $z/t$. Applying a general recurrence relation for the

Bessel function, $J_{m+1}(x) = J_{m-1}(x) - 2J'_m(x)$ and performing integration by parts on the derivative term, we find that,

$$G_{m+1} = -G_{m-1} + 2(z/t)G_m \qquad (42)$$

where $F_m = e^{i\psi}G_m$. This is exactly the recursion relation satisfied by (38).

The next is a proof that $\bar{p}^*_{-m}(-\vec{k}) = \bar{p}_m(\vec{k})$ for the case in which the absorption is real, which is useful for checking code. Here, $\bar{p}_m(\vec{k})$ is the FT of the weighted projection, as in Eq. 36, copied here:

$$\bar{p}_n(\vec{k}) = \sum_{m=-2}^{2} F_{n+m}(\vec{k})\bar{\mu}_m \qquad (43)$$

To derive this, we start with the original form for the weighted projections (eq. 28)

$$p_n(\vec{x}) = \sum_{m=-2}^{2} \int_0^{2\pi} d\gamma \exp(i(n+m)\gamma) \int dl\, \mu_m(\vec{x} + l\hat{n}(\gamma)) \qquad (44)$$

and take its Fourier transform:

$$\bar{p}_n(\vec{k}) = \frac{1}{(2\pi)^3} \int d^3\vec{x}\, e^{i\vec{k}\vec{x}} \sum_{m=-2}^{2} \int_0^{2\pi} d\gamma \exp(i(n+m)\gamma) \int dl\, \mu_m(\vec{x} + l\hat{n}(\gamma)) \qquad (45)$$

Now take the complex conjugate and substitute $-n$ for $n$:

$$\bar{p}^*_{-n}(\vec{k}) = \frac{1}{(2\pi)^3} \int d^3\vec{x}\, e^{-i\vec{k}\vec{x}} \sum_{m=-2}^{2} \int_0^{2\pi} d\gamma \exp(i(n-m)\gamma) \int dl\, \mu^*_m(\vec{x} + l\hat{n}(\gamma)) \qquad (46)$$

Next, change $\vec{k}$ for $-\vec{k}$ and use the identity $\mu^*_m = \mu_{-m}$:

$$\bar{p}^*_{-n}(-\vec{k}) = \frac{1}{(2\pi)^3} \int d^3\vec{x}\, e^{i\vec{k}\vec{x}} \sum_{m=-2}^{2} \int_0^{2\pi} d\gamma \exp(i(n-m)\gamma) \int dl\, \mu_{-m}(\vec{x} + l\hat{n}(\gamma)) \qquad (47)$$

and finally substitute $m$ for $-m$ in the sum:

$$\bar{p}^*_{-n}(-\vec{k}) = \frac{1}{(2\pi)^3} \int d^3\vec{x}\, e^{i\vec{k}\vec{x}} \sum_{m=-2}^{2} \int_0^{2\pi} d\gamma \exp(i(n+m)\gamma) \int dl\, \mu_m(\vec{x} + l\hat{n}(\gamma)) = \bar{p}_n(\vec{k})$$

$$(48)$$

We can use this result to derive a symmetry condition for the matrix that relates the observables to the unknowns. The piled-up system of equations can be written as

$$\bar{P}_{cm}(\vec{k}) = T_{cmq}(\vec{k})\bar{Q}_q(\vec{k}) \qquad (49)$$

with $T_{cmq}$ being the piled-up matrices in Eq. 41, $c$ an index to 'conditions' (tilt and polarization) and $q$ the index to the component of the tensor absorption. The repeated index is summed over, and $\bar{Q}_q$ is the FT of the $q$th tensor component. Again, assuming the tensor to be real, we have $\bar{Q}^*(\vec{k}) = \bar{Q}(-\vec{k})$. Taking the complex conjugate of (49), we have

$$\bar{P}^*_{cm}(\vec{k}) = T^*_{cmq}(\vec{k})\bar{Q}^*_q(\vec{k}) = T^*_{cmq}(\vec{k})\bar{Q}_q(-\vec{k}) \qquad (50)$$

Using the previous identity $\bar{p}^*_{-m}(-\vec{k}) = \bar{p}_m(\vec{k})$ we have

$$\bar{P}_{c-m}(-\vec{k}) = T^*_{cmq}(\vec{k})\bar{Q}_q(-\vec{k}) = T_{c-mq}(-\vec{k})\bar{Q}_q(-\vec{k}) \qquad (51)$$

with the last equality coming from (49) with the substitutions $\vec{k} \to -\vec{k}; m \to -m$. Thus, we have

$$T^*_{cmq}(\vec{k}) = T_{c-mq}(-\vec{k}) \quad . \tag{52}$$

A nice feature of this result is that it's very general and doesn't depend on the correctness of the math used to find an explicit form for $T_{cmq}$. It should be valid for MCD as well, replacing $Q$ with a 4-component object $(\mu_{nd}, \vec{m})$ with the variables as in Eq. 10 in the paper,

$$\mu(\vec{x}, \gamma, s) = \mu_{nd}(\vec{x}) + s\vec{m}(x) \cdot \hat{n}(\gamma) \quad . \tag{53}$$

The form given in the paper has a phase rotation $e^{i\psi m}$ where $\psi$ is the polar angle of the point in k-space. In coding, it can be hard to make sure that this is applied correctly.

Now let's dissect $T$ into its component parts, $G$ and $S$ and look for similar symmetry conditions. We can express the $S$ matrix as:

$$S_{mq} = \frac{1}{2\pi} \int_0^{2\pi} d\gamma \exp(-im\gamma) \mu_q(\gamma) \tag{54}$$

where $\mu_q$ is the contribution to the absorption due to a unit value of $Q_q$ with no other element non-vanishing. This is the form I used to derive the form for the $S$-matrix shown in Eq. 23 in the paper. It is obvious, then, that

$$S_{-mq} = S^*_{mq} \tag{55}$$

with the assumption that the absorption tensor is real. Now, given $T_{cmq}(\vec{k}) = G_{mm'}(\vec{k}) S_{m'q}$ and taking the complex conjugate, we get

$$T^*_{cmq}(\vec{k}) = T_{c-mq}(-\vec{k}) = G^*_{mm'}(\vec{k}) S^*_{m'q} = G^*_{mm'}(\vec{k}) S_{-m'q} = G^*_{m-m'}(\vec{k}) S_{m'q}$$

$$G^*_{m-m'}(\vec{k}) = G_{c-mm'}(-\vec{k}) \tag{56}$$

$$G^*_{mm'}(\vec{k}) = G_{c-m-m'}(-\vec{k})$$

Thus, we see that $G$ becomes its complex conjugate on negation of both of its indices and its $k$-point argument.

## 7. Reconstruction program notes

### 2.1 Introduction

This note provides some of the details on the reconstruction algorithm discussed in main text and the simulated tomography code used to demonstrate the algorithm.

There are three parts to the system:
1. Generation of a dichroic phantom and simulation of tomographs
2. Generation of the matrix used in Eq. (41) in main text, which is the design matrix of a least-squares problem solved to perform the reconstruction
3. The actual reconstruction

In addition, there are details in common to two or more parts.

### 2.2 Commonalities

A 'condition' is defined as a tilt-angle/polarization-angle pair $(\alpha, \chi)$. In order to give ASTRA the information it needs to perform projections and back-projections, several vectors need to be defined. These vectors are expressed in the sample's coordinates, in which the $z$ axis is the rotation axis. Most of these are given in the main text, but for reference, they are:

Ray direction ($\hat{n}$, Eq. (8)): $(\sin\alpha\cos\gamma, -\sin\alpha\sin\gamma, \cos\alpha)$

Detector center position: (0,0,0)

Vector from detector pixel (0,0) to (0,1) ($\hat{Y}$ in sample coords): $(\sin\gamma, \cos\gamma, 0)$

Vector from detector pixel (0,0) to (1,0) ($\hat{X}$ in sample coords):
$(-\cos\alpha\cos\gamma, \cos\alpha\sin\gamma, \sin\alpha)$

Polarization vector ($\hat{e}$ in sample coords, Eq. (9)):
$(\sin\gamma\cos\chi - \cos\gamma\cos\alpha\sin\chi, \cos\gamma\cos\chi + \sin\gamma\cos\alpha\sin\chi, \sin\alpha\sin\gamma)$

The operation of 'flattening' two or more indices into one is used two ways. In one, the condition and $m$ (moment) indices are flattened so that the weighted back-projection belonging to condition $c$ and moment $m$ can be represented as $p_{c,m}(\vec{x}) \rightarrow p_{cm}(\vec{x})$. Also, Fourier transforms, which are three-dimensional and so have three indices, are flattened to 1D. It is necessary to exclude from calculation Fourier coefficients inside the missing cone. Thus, the Fourier transformed weighted back-projections, which in full-index notation would be $\tilde{P}(k_x, k_y, k_z; c, m)$ is instead $\tilde{P}(ik, cm)$ and the matrix equations to be solved (in a least-squares sense) are $T_{ik,cm,q}\overline{Q}_{ik,q} = \tilde{P}_{ik,cm}$, with one such equation for each allowed $k$-point. Here $T$ is the big, piled matrix in Eq. (41) and $\overline{Q}_q$ is the Fourier transform of the 6-element vector $Q$, which is the quantity to be solved for. After the solution, the flattened form is unpacked into three dimensions, with Fourier components in the missing cone set to 0.

## 2.3 Generation

The phantom is defined as a six-component vector Q[z,x,y,q] where z,x,y specify the voxel (the order is what ASTRA uses) and q is the index to the tensor component as in Eq. (21). The volume is 200x200x200 voxels and the phantom is non-zero within a cube 25 voxels across. These tensor components are written out into .tif stacks for comparison with the reconstruction. Then, to create the projections, the ASTRA routines data3d.create() and creators.create_sino3d_gpu() are used once for each γ angle, with the absorption coefficient at each voxel given by
$$\mu(z,x,y;\gamma) = \hat{e}_x^2 Q_1(z,x,y) + \hat{e}_y^2 Q_2(z,x,y) + \hat{e}_z^2 Q_3(z,x,y) + \hat{e}_x\hat{e}_y Q_4(z,x,y) + \hat{e}_x\hat{e}_z Q_5(z,x,y) + \hat{e}_y\hat{e}_z Q_6(z,x,y)$$
. These projections constitute the simulated data. This operation is done for each condition. The first two steps of reconstruction, making the weighted back-projections and writing out their Fourier transforms, are done in the same program that does the simulation.

Next, the projections are multiplied by $\cos m\gamma$ and $\sin m\gamma$, $m \in [-2, 2]$ (real and imaginary parts of $\exp im\gamma$-weighting) and back-projected using the ASTRA routines data3d.create() and create_backprojection3d_cuda(). An important but non-obvious step is that the back-projections need to be 'tapered' so as to go to 0 at the faces of the data cube. This is done by multiplying by a function $t(z, y, x) = t_1(x)t_1(y)t_1(z)$ with $t_1(x) = 4x(1-x)$ and the coordinates scaled to go between 0 and 1. This function is 1 in the center and 0 at the edges. Without this step, the reconstructions are unrecognizable. It is possible that the problem is that the data are non 0-padded before Fourier transforming, which leads to discontinuities at the faces of the data cube. Finally, the Fourier transforms are taken and written out as `.tif` stacks. Since the Fourier transforms are complex, the real and imaginary parts are written out separately.

## 2.4 The T-matrix

The big matrix could be generated analytically, using the equations in the paper. However, it's simpler to go through the steps outlined above for generation of simulated data, including the taper, but with the phantom having a vanishing $Q$ tensor except for one point in the center of the volume, and then Fourier transforming the data. This procedure is carried out six times, with one component of $Q$ being set to 1 at the center voxel. A phase factor is applied to the Fourier transforms because the center voxel is not at (0,0,0) but at $(n/2, n/2, n/2)$ with $n$ the size of the cube. The Fourier transforms are written out as `.tif` stacks, again separated into real and imaginary parts. These are the elements of $T_{k_z, k_x, k_y, icm, q}$. For a set of four conditions, there are 4x5x2x6=240 such files, with the factor of 2 being due to the need to write real and imaginary parts. The matrix generation need only be carried out anew if the conditions are changed; the same matrix applies to any data to be reconstructed as long as it was generated under the same conditions as the matrix and is of the same size in all directions.

In principle, this direct-generation method for evaluating the matrix may compensate to some extent for missing angles if the point "phantom" is "imaged" with the same set of angles as in the actual data. This has not been tested.

## 2.5 Reconstruction

The reconstruction program reads the Fourier-transformed back-projections produced by the generation program and the matrix element stacks produced by the $T$-matrix program. It then performs the flattening in $k$-space, followed by least-squares solution and then un-flattening. The matrix condition number and rank are evaluated at a sampling of $k$-points. In the case investigated, the rank is always 6, as desired, and the condition number ranges from 5 to 50. The un-flattened Fourier components are then processed by an inverse Fourier transform, resulting in the reconstructed real-space data, which is written into `.tif` stacks. The output of the inverse Fourier transform is complex, but if the reconstruction worked properly the imaginary part should be zero. Thus, looking at the imaginary parts of the reconstructed components provides a basic sanity check.

**References**

1.  P. Fischer, "X-ray imaging of magnetic structures," IEEE Transactions on Magnetics **51**, 1-31 (2014).
2.  J. Stöhr and H. C. Siegmann, "Magnetism," Solid-State Sciences. Springer, Berlin, Heidelberg **5**(2006).
3.  Y.-S. Yu, R. Celestre, B. Enders, K. Nowrouzi, H. Padmore, T. Warwick, J.-R. Jeong, and D. A. Shapiro, "Nanoscale Visualization of Magnetic Contrasts with Soft X-ray Spectro-Ptychography at the Advanced Light Source," Microscopy and Microanalysis **24**, 530-531 (2018).
4.  K. G. Rana, R. L. Seeger, S. Ruiz-Gómez, R. Juge, Q. Zhang, V. T. Pham, M. Belmeguenai, S. Auffret, M. Foerster, and L. Aballe, "Room temperature skyrmions in an exchange biased antiferromagnet," arXiv preprint arXiv:2009.14796 (2020).
5.  C. Donnelly, M. Guizar-Sicairos, V. Scagnoli, S. Gliga, M. Holler, J. Raabe, and L. J. Heyderman, "Three-dimensional magnetization structures revealed with X-ray vector nanotomography," Nature **547**, 328-331 (2017).
6.  H. Braun and A. Hauck, "Tomographic reconstruction of vector fields," IEEE Transactions on signal processing **39**, 464-471 (1991).
7.  C. A. Stifler, J. E. Jakes, J. D. North, D. R. Green, J. C. Weaver, and P. U. Gilbert, "Crystal misorientation correlates with hardness in tooth enamels," Acta Biomaterialia **120**, 124-134 (2021).

8. R. T. DeVol, R. A. Metzler, L. Kabalah-Amitai, B. Pokroy, Y. Politi, A. Gal, L. Addadi, S. Weiner, A. Fernandez-Martinez, R. Demichelis, J. D. Gale, J. Ihli, F. C. Meldrum, A. Z. Blonsky, C. E. Killian, C. B. Salling, A. T. Young, M. A. Marcus, A. Scholl, A. Doran, C. Jenkins, H. A. Bechtel, and P. U. P. A. Gilbert, "Oxygen spectroscopy and Polarization-dependent Imaging Contrast (PIC)-mapping of calcium carbonate minerals and biominerals," J. Phys. Chem. B **118**, 8449-8457 (2014).
9. H. Chang, M. A. Marcus, and S. Marchesini, "Analyzer-free linear dichroic ptychography," Journal of Applied Crystallography **53**(2020).
10. Y. H. Lo, J. Zhou, A. Rana, D. Morrill, C. Gentry, B. Enders, Y.-S. Yu, C.-Y. Sun, D. A. Shapiro, and R. W. Falcone, "X-ray linear dichroic ptychography," Proceedings of the National Academy of Sciences **118**(2021).
11. Z. Gao, M. Holler, M. Odstrcil, A. Menzel, M. Guizar-Sicairos, and J. Ihli, "Nanoscale crystal grain characterization via linear polarization X-ray ptychography," Chemical Communications **56**, 13373-13376 (2020).
12. J. Lüning, F. Nolting, A. Scholl, H. Ohldag, J. W. Seo, J. Fompeyrine, J.-P. Locquet, and J. Stöhr, "Determination of the antiferromagnetic spin axis in epitaxial LaFeO 3 films by x-ray magnetic linear dichroism spectroscopy," Physical Review B **67**, 214433 (2003).
13. D. Zhao, H. Höchst, and D. L. Huber, "Dielectric tensor formulation of magnetic dichroism sum rules," Journal of applied physics **84**, 2858-2860 (1998).
14. F. Pfeiffer, "X-ray ptychography," Nature Photonics **12**, 9-17 (2018).
15. A. van der Sluis and H. A. van der Vorst, "Numerical solution of large, sparse linear algebraic systems arising from tomographic problems," in *Seismic Tomography: With Applications in Global Seismology and Exploration Geophysics*, G. Nolet, ed. (Springer Netherlands, Dordrecht, 1987), pp. 49-83.
16. I. Wolfram Research, "Mathematica" (2021), retrieved https://www.wolfram.com/mathematica.
17. A. Myagotin, A. Voropaev, L. Helfen, D. Hänschke, and T. Baumbach, "Efficient volume reconstruction for parallel-beam computed laminography by filtered backprojection on multi-core clusters," IEEE Transactions on Image Processing **22**, 5348-5361 (2013).
18. H. Matsuo, A. Iwata, I. Horiba, and N. Suzumura, "Three-dimensional image reconstruction by digital tomo-synthesis using inverse filtering," IEEE transactions on medical imaging **12**, 307-313 (1993).
19. S. Fisher, D. Holmes, J. Jørgensen, P. Gajjar, J. Behnsen, W. Lionheart, and P. Withers, "Laminography in the lab: imaging planar objects using a conventional x-ray CT scanner," Measurement Science and Technology **30**, 035401 (2019).
20. S. Harasse, W. Yashiro, and A. Momose, "Iterative reconstruction in x-ray computed laminography from differential phase measurements," Optics Express **19**, 16560-16573 (2011).
21. W. Van Aarle, W. J. Palenstijn, J. Cant, E. Janssens, F. Bleichrodt, A. Dabravolski, J. De Beenhouwer, K. J. Batenburg, and J. Sijbers, "Fast and flexible X-ray tomography using the ASTRA toolbox," Optics express **24**, 25129-25147 (2016).
22. C. Donnelly, S. Gliga, V. Scagnoli, M. Holler, J. Raabe, L. J. Heyderman, and M. Guizar-Sicairos, "Tomographic reconstruction of a three-dimensional magnetization vector field," New Journal of Physics **20**, 083009 (2018).
23. A. Hierro-Rodriguez, C. Quirós, A. Sorrentino, L. M. Álvarez-Prado, J. Martín, J. M. Alameda, S. McVitie, E. Pereiro, M. Velez, and S. Ferrer, "Revealing 3D magnetization of thin films with soft X-ray tomography: magnetic singularities and topological charges," Nature communications **11**, 1-8 (2020).
24. A. Hierro-Rodriguez, D. Gürsoy, C. Phatak, C. Quirós, A. Sorrentino, L. M. Álvarez-Prado, M. Vélez, J. I. Martín, J. M. Alameda, and E. Pereiro, "3D reconstruction of magnetization from dichroic soft X-ray transmission tomography," Journal of synchrotron radiation **25**, 1144-1152 (2018).
25. D. M. Pelt, K. J. Batenburg, and J. A. Sethian, "Improving tomographic reconstruction from limited data using mixed-scale dense convolutional neural networks," Journal of Imaging **4**, 128 (2018).
26. M. Abromowitz and I. A. Stegun, "Handbook of mathematical functions," NBS (now NIST) (1965).